\documentstyle[epsfig]{aipproc}

\begin{document}
\def\vp{\vec p\,}
\def\v0{\vec 0\,}
\def\beq{\begin{eqnarray}}
\def\eeq{\end{eqnarray}}

\title{A CP-Violating Kinematic Structure}

\author{D. V. Ahluwalia}

\address{Escuela de Fisica, ISGBG, Ap. Postal C-600\\
Univ. Aut. de Zacatecas\\
Zacatecas, ZAC 98068, Mexico}

\maketitle

\begin{abstract}
A CP violating kinematic structure is presented. The essential physical
input is to question the textbook wisdom, ``Now when a particle is at 
rest, one cannot define its spin as either left- or right-handed, 
so $\phi_R(0) = \phi_L(0)$,'' as found, e.g., in Lewis Ryder's 
{\em Quantum Field  Theory,\/} and in many other books on the representations
of the Lorentz group. It is suggested that this equality is true only
up to a phase. The demand of  C, P, and T covariances, separately, 
fixes this phase to be $\pm 1$. 
If these conditions are relaxed, 
a natural CP-violating kinematic structure emerges.
Having established a CP-violating kinematic structure,
we then discuss how Planck scale physics necessarily invokes non-commutative
space-time  and that such changes in the structure of space-time will force
upon us additional violations/deformations of the CPT structure of space-time, 
and a violation of the principle of equivalence via a violation
of the Lorentz symmetries. The latter may carry
significant consequences for understanding the data on ultra high energy
cosmic rays.
\end{abstract}

\section*{Introduction}
The moment one invokes the Poincar\'e 
space-time symmetries  the notions of {\em mass\/},  
{\em spin}, and the existence of  {\em two types of matter,\/} i.e. 
particles and antiparticles, immediately arise. 
Poincar\'e symmetries also play equally important role in the
gravitational realm.   
One can even argue that the standard wave-particle duality 
carries its basis in the generators of space-time translations. The
eigenstates of these generators, which can be superimposed to make
normalizable physical states, carry a spatial
periodicity that, in the presence of a non-vanishing Planck constant,
can be identified with the de Broglie wave length.

Here, I argue that the quantum mechanical framework induces new
elements in  the C, P, and T structure of the representations offered by 
the Lorentz group. Arguments of this nature were long ago
appreciated by G\"ursey, Michel, Wigner,
and by other physicists of their generation, see, e.g., Ref. \cite{fg} and 
pp. 453-457 of Ref. \cite{Vigier}, and also refer to the Lee and Wick paper 
\cite{lw}. The new element that I mentioned in the {\em abstract}
turns out, in a very precise sense \cite{bww}, 
to be a missing link that prevented
construction of the Wigner classes (see Wigner in Ref. \cite{fg}).
These arguments provide unsuspected source of CP violation. 
It emerges, e.g., that the violation 
of CP may also be found in a new quantum-induced
CP-violating kinematic structure offered by the 
Lorentz group, rather than in a new interaction. To clarify, one should 
note that in the standard gauge theory of the electroweak interactions 
the underlying kinematic structure manifestly violates parity. 
The gauge group of the weak interactions serves 
to provide the gauge bosons that mediate the interaction on 
this P-violating kinematic structure. The credit on gauge bosons, 
and the Higgs boson, is to help set the range of this P-violating 
interaction in a renormalizable theory. 
Our general results are not in conflict with the common 
wisdom, see, e.g., Ref. 
\cite{ew}, because these works confine to the usual kinematic structure 
for the spin-1/2 fermions and spin-1 vector bosons (note, just because
a particle carries spin one does not necessarily make it a vector object)
and which seek CP violation in interactions.
Purely on the grounds of representations of the Lorentz group in 
the quantum realm, we shall discover the possibility 
of a new phase field (called  $\Phi(\theta)$ below). In this talk we 
shall remain far from identifying the (scalar) phase field with 
the Higgs, but the existence of the new field arises so naturally 
that there may be an important place for it in physics.

This also sets the stage to study CPT-related consequences of the possible
deformations of the space-time symmetries
required by the joint realm of the 
quantum and gravity.
On the empirical side, the data on the ultra high energy cosmic 
rays has taken us to within  
nine orders of magnitude of the Planck scale.
That data has already raised questions that invoke the violation 
of the Lorentz symmetries. In this context too, I find 
the reflections presented here to be more than of academic interest.

\section*{The Quantum-Induced CPT Structure of Space-time}

The kinematic structure of the existing quantum theory of fields
originates from the Poincar\'e symmetries \cite{epw1939}. 
This is well explained in  recent books of Ryder, Sterman, 
and Weinberg \cite{lhr,gs,sw}. For the purpose at hand the reader 
is also referred to my book reviews \cite{r1,r2}, and Ref. 
\cite{gg}. In these references certain conceptual issues 
have been clarified and corrected. 


I now provide a brief, but careful, review of the spin-$1/2$ 
representation space.
If we  wish to construct
a parity covariant  spin-$1/2$ representation space then the 
fields operators associated with the kinematic structure 
must be constructed in the $(1/2,0)\oplus(0,1/2)$ representation space.
One must further decide whether one wishes to describe charged particles
in the Dirac sense,
or neutral  particles in the Majorana sense. 
To treat both of these constructs at
an equal footing the underlying spinors must carry similar
C, P, and T properties as the field operators themselves. For the
usual Majorana field operator the just-stated requirement is 
badly violated because it is expanded in terms of the {\em Dirac} spinors.
In the usual textbook constructs, see, e.g., Ref.
\cite{rm}, it is only in terms of the Fock space creation and annihilation 
operators that the distinction is made between the Dirac and Majorana
field operators.  In these constructs, we repeat,
the underlying spinors, for the Majorana field
operator, are still the Dirac spinors.

To describe the fundamentally {\em charged\/} particles
in the Dirac sense the appropriate $(1/2,0)\oplus(0,1/2)$ 
spinors have the form:
\beq
\psi(\vp) \equiv \left(
			\begin{array}{c}
			\phi_R(\vp)\\
                        \phi_L(\vp)
                        \end{array}
                   \right).
\eeq
The $\phi_R(\vp)$ transforms as a $(1/2,0)$ spinor,\footnote{
Note, identical transformation properties under Lorentz group
do not necessarily imply that other transformations properties
will be identical as well. The latter, e.g., may refer to transformations
under C, P, and T. Thus, the Dirac- and Majorana-$(1/2,0)\oplus(0,1/2)$
constructs carry different physical characteristics under operations
of C, P, and T, while carrying identical transformations 
under the Lorentz group.}
and boosts as 
\beq
\phi_R(\vp) = \exp\left(\frac{\vec\sigma}{2}\cdot\vec\varphi\right)
\phi_R(\v0).\label{r}
\eeq
In the above equation, the $\vec \sigma$ are the usual Pauli matrices, and
$\vec\varphi$ is the boost parameter. The definition of $\vec\varphi$
is motivated by the fact that $E^2-
\vp^2 = m^2$, and that $\cosh^2\alpha -\sinh^2\alpha=1$ (as an
identity). Thus, 
\beq
\cosh\varphi=\frac{E}{m}, \quad \sinh\varphi=\frac{\vert\vp\vert}{m}.
\label{bp}
\eeq  
The direction associated with $\vec \varphi$ is that of the three
momentum associated with the particle: 
\beq
\widehat\varphi = \frac{\vec p}{\vert\vp\vert}.
\eeq
We shall assume that $m\ne 0$.

Whereas, the $\phi_L(\vp)$ transforms is a $(0,1/2)$ spinor, and boosts
with the opposite sign in the exponent:  
\beq
\phi_L(\vp) = \exp\left(- \frac{\vec\sigma}{2}\cdot\vec\varphi\right)
\phi_L(\v0).\label{l}
\eeq
The $\v0$ corresponds to the momentum vector for the 
particle at rest.

\vskip 1cm

On the other hand, if one wishes to describe fundamentally {\em 
neutral\/} particles in the Majorana sense 
then the correct choice of the  $(1/2,0)\oplus(0,1/2)$ spinors is:
\beq
\lambda(\vp) = \left(\begin{array}{c}
                        \zeta_\lambda \Theta_{[1/2]} \phi^\ast_L(\vp)\\
                        \phi_L(\vp)
                        \end{array}
		\right),\label{lambda}
\eeq
and
\beq
\rho(\vp) =  \left(\begin{array}{c}
                \phi_R(\vp)\\
                \left(\zeta_\rho\Theta_{[1/2]}\right)^\ast \phi^\ast_R(\vp)
	\end{array}
	\right).\label{rho}
\eeq
In these expressions $\Theta_{[1/2]}$ is the spin-$1/2$ 
Wigner time reversal operator:
\beq
\Theta_{[1/2]} = \left(\begin{array}{cc}
                        0 &-1\\
                        1 & 0
			\end{array}
	\right).
\eeq
The phases factors $\zeta_\lambda$ and $\zeta_\rho$, that appear
in Eqs. (\ref{lambda}) and (\ref{rho}), are determined by the
condition of self(S)/anti-self(A) conjugacy under the operation of 
charge conjugation. In Refs. \cite{dva1996,agj} it was shown that this
condition determines these phases to be:
\beq
\zeta^S_\lambda &=& + i =\rho^S_\rho,\\
\zeta^A_\lambda &=& - i =\rho^A_\rho,
\eeq
Incorporating the
anti-self conjugate spinors is essential for the complete, in the
sense of mathematically complete set, description of the neutral
$(1/2,0)\oplus(0,1/2)$ representation space. The classic 1957
McLennan-Case reformulation of the Majorana theory was incomplete
in this aspect. Similarly, there is a wide spread, but incorrect, belief that 
somehow Majorana neutral objects carry half as many degrees
of freedom as Dirac neutral objects.  
The reader who wishes to scrutinize these issues is
referred to Ref. \cite{dva1996}.

\vskip 1cm
The {\em wave equations\/} satisfied by these spinors follow from: 
(a) The transformation properties of the $(1/2,0)$ and $(0,1/2)$ 
spinors, and
very importantly, (b) The relative phase between the $(1/2,0)$ and 
$(0,1/2)$ spinors at rest. Because this may carry important physical
consequences we establish this claim. We shall first carefully examine
the $(1/2,0)\oplus(0,1/2)$ representation space suitable for describing
{\em charged\/} particles. Only later shall we return to the {\em 
neutral\/} particles very briefly.

Due to the isotropy of the $\vec p = \vec 0$, one may argue that
$\phi_R(\v0) = \phi_L(\v0)$. In fact that is precisely what is
done in the standard textbooks \cite{lhr,Hladik}. 
Ryder's classic book on the theory of quantum fields \cite{lhr},
in fact, argues, ``Now when a particle is at rest, one cannot define 
its spin as either left- or right-handed, so $\phi_R(0) = \phi_L(0)$.'' 
This may be argued to some extent if one was to
confine to a purely classical framework. In the process, as it turns
out, what one misses are anti-particles!
However, if one is
to invoke a quantum framework for the interpretation of these spinors
then this equality can only be claimed up to a phase:
\beq
\phi_R(\v0) = e^{i\theta} \phi_L(\v0).\label{rl}
\eeq
This is the central physical input which will yield us the result
(\ref{eq}). In fact, in a private communication, Ryder has pointed
out that $\theta$ may be a $2\times 2$ matrix. Here, we shall confine
to the simplest suggestion that $\theta$ is a real angle (possibly
carrying a space-time dependence).

Equations (\ref{r}), (\ref{l}), and (\ref{rl})  contain 
essentially\footnote{We qualify with ``essentially'' because while
constructing the field operators the Fock space
considerations must be invoked, in addition.}
the entire kinematic structure 
of the {\em charged\/} spin-$1/2$ particles. To see
this we follow the foot steps of Lewis Ryder \cite{lhr}, 
but we now carefully incorporate the important ingredient embedded in a 
non-vanishing $\theta$.  

\begin{quote}
\begin{enumerate}
\item

On the right-hand side of Eq. (\ref{r}), substitute for $\phi_R(\v0)$
from (\ref{rl}). This gives
\beq
\phi_R(\vp)=e^{i\theta} \exp\left(\frac{\vec\sigma}{2}\cdot\vec\varphi\right)
\phi_L(\v0). \label{a}
\eeq

\item
From Eq. (\ref{l}) obtain, 
\beq
\phi_L(\v0)=
\exp\left(\frac{\vec\sigma}{2}\cdot\vec\varphi\right)\phi_L(\vp),
\eeq
and insert it into the right-hand side of 
Eq. (\ref{a}). This yields:
\beq
\phi_R(\vp)=e^{i\theta}\exp\left(\vec\sigma\cdot\vec\varphi\right)
\phi_L(\vp).\label{b}
\eeq

\item
Similarly, starting from Eqs. (\ref{l}) and (\ref{rl}) we obtain:
 \beq
\phi_L(\vp)=e^{-i\theta}\exp\left(-\vec\sigma\cdot\vec\varphi\right)
\phi_R(\vp).\label{c}
\eeq

\item
Now, because $
\left(\vec\sigma\cdot\widehat{p}\right)^2 = 2\times 2\,\,\mbox{Identity
matrix,}\,\, I_2$
\beq
\left(\vec\sigma\cdot\widehat{p}\right)^n=\cases{I_2 & for n even\cr
                                \vec\sigma\cdot\widehat{p} & for n odd\cr}.                 
 \label{Pauli}
\eeq
This leads to the identities:  
\beq
\exp\left(\pm \vec\sigma\cdot\vec\varphi\right)
=\frac{E I_2\pm\vec\sigma\cdot\vec p}{m}
\eeq

\item
Next, substitute these identities in Eqs. (\ref{b}) and (\ref{c}), 
and re-arrange to obtain:
\beq
\left(
\begin{array}{cc}
-m e^{-i\theta} & E I_2 +  \vec\sigma\cdot\vec p \\
E I_2- \vec\sigma\cdot\vec p & -m e^{i\theta}
\end{array}
\right)
\left(
\begin{array}{c}
\phi_R(\vp)\\
\phi_L(\vp)
\end{array}
\right)= 0.
\eeq
\item
Finally, with $p_\mu=(p^0, \,-\vp)$, $E=P^0$, 
read off the Weyl-representation
gamma matrices:\footnote{We now abbreviate 
$I_2$ by $I$. The zeros below stand for
$2\times 2$ null matrices, and $\sigma^1=\sigma_x$, etc.}
\beq
\gamma^0= 
\left(
\begin{array}{cc}
0 & I\\
I & 0
\end{array}
\right), \quad 
\gamma^i= 
\left(
\begin{array}{cc}
0 & - \sigma^i\\
\sigma^i & 0
\end{array}
\right),
\eeq

and introduce
\beq
\Phi(\theta)= 
\left(
\begin{array}{cc}
\exp(-i\theta) & 0\\
0 & \exp(+i\theta)
\end{array}
\right), 
\eeq
This yields, the central result of our thesis:
\beq
\Big(\gamma^\mu p_\mu - m\Phi(\theta)\Big)\psi(\vec p) =0.
\label{eq}
\eeq

\end{enumerate}
\end{quote}

The obtained equation is Poincar\'e covariant and indeed carries
the solutions with the correct dispersion relations $E=\pm\sqrt{\vp^2+m^2}$,
because not only 
\beq
\mbox{Det}\left[\gamma^\mu p_\mu - m\Phi(\theta)\right]
=\left(\vp^2+m^2-E^2\right)^2
\eeq
but also because 
$\mbox{Det}\left[\gamma^\mu p_\mu - m\Phi(\theta)\right]$
is independent of $\theta$. Thus, as expected, Poincar\'e covariance
cannot constrain $\Phi(\theta)$. The $\Phi(\theta)$ is constrained to
be $\pm 1$ if one places the extra condition that the resulting equation
be covariant {\em separately} under operations of C, P, and T. The
$\Phi(\theta)$s cannot be ``rotated away'' particularly 
if one considers a general system of more than one spin-$1/2$. 
Further, in absence of the indicated CPT-related
covariances, $\Phi(\theta)$ may carry a space-time dependence. Note, without
the  $\Phi(\theta) = -1$, the ``antiparticle'' solutions are missed as was
brought to the reader's attention earlier.

We thus conjecture that
$\Phi(\theta)$ may carry space-time dependence (particularly,
in curved space-time) --- and even if it were
to be taken as a constant matrix --- $\Phi(\theta)$ carries information
on the CPT structure of the spin-1/2 {\em charged} fields. 
It is too
premature to speculate if  $\Phi(\theta)$ may not be related to the
observed CP violation in the universe.
However, if one insists on a {\em renormalizable\/} theory with
{\em no space-time dependence\/} in $\Phi(\theta)$, then considerations
of Weinberg, see Sec. 12.5 of Ref. \cite{sw}, could render $\Phi(\theta)$
physically unobservable. The CP violating kinematic structure obtained
here is also contained in the postulated ``CP violating Dirac equation''
of Funakubo {\em et al.\/} \cite{f}. For {\em non local\/} elements in this
structure, and additional kinematic details, the reader is referred
to Ref. \cite{cp}.

Before proceeding further, let's take note that the existence of
$\Phi(\theta) = -1$, missed in the well-known classical arguments, and 
in fact not allowed by the realm of classical framework, is a 
consequence of the quantum mechanical freedom encoded in Eq. (\ref{rl}).
For a C--, P--, and T--covariant theory, existence of {\em two values,\/} 
rather than one, for $\Phi(\theta)$ is responsible for the existence
of two types of matter, i.e. particles and antiparticles. In the standard
treatment of spin-1/2 particles, the quantum-mechanical origin of this 
fact was well-understood by Dirac and other knowledgeable physicists, 
even though by no means existence of the phase field $\Phi(\theta)$
was so transparent, or was even suspected. Second, the linearity of Eq. 
(\ref{eq}) in $p_\mu$ is a consequence of the very specific 
property of the Pauli matrices contained in Eq. (\ref{Pauli}). Third, 
this linearity is not guaranteed if one considers  {\em neutral\/}
particles in the Majorana sense, see Ref. \cite{dva1996}. 

When one repeats the same exercise for the {\em charged} particles
in the $(1,0)\oplus(0,1)$ representation space one obtains a new, and
to an extent unexpected, result that such spin-1 particles, as opposed
to spin-1 vector particles, are of a Wigner class which was presented
only relatively recently in Ref. \cite{bww}. 
In this Wigner class, a boson and its
antiboson carry {\em opposite\/} relative intrinsic parity. This arises
because the C and P operators of the $(1,0)\oplus(0,1)$ representation
space {\em anticommute}, rather than commute. 
No search for such particles
in the low energy domain has yet been undertaken. Because the natural
counterpart of the spin-$1/2$ $\gamma^\mu$ in the $(1,0)\oplus(0,1)$ 
representation space is a set of two-indexed 
$6\times 6$ matrix objects $\gamma^{\mu\nu}$ they may provide a natural
coupling to the space-time metric $g_{\mu\nu}$. For this reason, these
particles may have played an important role in the early universe. 
In particular, their unexpected properties as regards their C, P, and T 
transformations make them natural candidates for studying any unusual 
CP properties of the early phase of the universe. Once again, a phase
field similar to $\Phi(\theta)$ above can be introduced for the 
$(1,0)\oplus(0,1)$ representation space.

Here we shall refrain from further discussing the representation spaces 
associated with the {\em neutral\/} particles. Once again, one finds
that neutral 
 $(1/2,0)\oplus(0,1/2)$ representation
space carries distinct, and different, structure under the transformations
of C, P, and T. The reader is referred to Refs. \cite{dva1996,agj,vd1,vd2}. 

\section*{Discussion: Possible Role of Gravity}

The CPT properties of the underlying kinematic structure
of the existing quantum field theories are thus explicitly seen to
have their entire roots in the space-time symmetries and the quantum 
mechanical  phase fields, such as $\Phi(\theta)$. A natural thesis
thus arises: The observed CP violation may carry
its origins in some new CP-violating kinematic structure rather than a
new interaction.

Having established these results it is now important to note that
if gravitational effects of quantum measurements are not neglected
then the resulting space-time is necessarily non-commutative, and
brings in certain elements of non-locality \cite{grf1994}. 
This happens because
the quantum mechanical collapse of a wave function also carries
with it an unavoidable collapse of the associated energy-momentum
tensor. This brings in an interplay of the gravitational
and quantum realms. Therefore, 
at the Planck scale one expects a space-time
that is non-commutative and which uses appropriate deformations of the
Poincar\'e group.  Such interplay of the quantum and gravitational
realms generically introduce 
gravitationally-modified wave-particle duality,\footnote{In this context
I refer the reader not only to the more physically motivated works, such
as \cite{grf1994,pla2000,ak,sdh,as,gac,mm,ns,fs,cls},
but also to more formal and important works represented by Refs. 
\cite{jm,ac,sm}.}
or equivalently in the context of string-theory suggested
modifications in the fundamental uncertainty relations \cite{string_ur}.
All this endows space-time with new quantum-induced modifications. 
This would have immediate consequences for the
underlying kinematic structure of the theory as regards its
CPT properties. This latter result,
on the one hand this seems suggested by the intrinsic non-locality
and non-commutative geometry
imposed by the interplay of the quantum and gravitational
realms \cite{grf1994}, and on the other by hand (which seems to contain
the same physical origin) by more formal studies devoted to investigating
non-commutative space-time. Therefore, it is  clear that that
deformations of the Poincar\'e symmetries, whether expected
on the grounds of the latest data on ultra high-energy cosmic
rays, or on the basis of 
more theoretical grounds, shall
have far reaching consequences for the entire structure on which
the present theoretical physics is based\cite{cr1,cr2,cr3,cr4,cr5}. 
In this context, it is
relevant to note that the instance Poincar\'e symmetries are
deformed one must expect a modification of the principle of 
equivalence also.

It is important to emphasize that a $\Phi(\theta)\ne \pm 1$ may lead
to low energy consequences, such as CP violation, whereas deformations
of the Poincar\'e symmetries and the associated change in the underlying
CPT structure of the kinematics shall be responsible for additional 
effects that would be dominant as we approach the Planck scale.

\vskip 1cm
\noindent
{\bf Acknowledgements}

First, my warm thanks to the organizers of this timely meeting 
in such a pleasant environment, and for inviting me to Metepec. 
Then, I want to acknowledge that I had three very particular 
readers in my mind when I started writing this talk. These are
Mariana Kirchbach, Yong Liu, and Lewis Ryder. To Mariana I extend
my thanks for an on-going discussion and for many many questions, 
to Lewis I send my thanks for he is the one whose book put me to 
all this in the first place, and to Yong 
I thank for being such a good student-teacher. I'm very proud 
of his work in \cite{yong} and for his uncompromised dedication
to physics in the ancient traditions of the East. 
Finally, my {\em zimpoic\/} thanks remain with Christoph Burgard
for many many useful discussions on the subject several years ago.

\def\bi{\bibitem}


\begin{references}

\bi{fg}
G\"ursey, F (editor), {\it Group Theoretical Concepts and Methods in Elementary
Particle Physics: Lectures of the Istanbul Summer School on 
Theoretical Physics,\/} New York: Gordon and Breach Science Publishers,
1964.

\bi{Vigier}
Ahluwalia, D. V. in Jeffers {\em et al.\/} (editors), 
{\it The Present Status of the Quantum Theory of Light: 
Proceedings of a Symposium in Honour of
Jean-Pierre Vigier,\/} Dordrecht: Kluwer Aacdemic Press, 1997.

\bi{lw}
Lee, T. D., and Wick, G. C., {\it Phys. Rev.\/} {\bf 148}, 1385 (1966).

\bi{bww}
Ahluwalia, D. V., Johnson, M. B., and Goldman, T.
{\em Phys. Lett. B\/} {\bf 316}, 102 (1993).

\bi{ew}
Einhorn, M. B., and Wudka, J., hep-ph/0007285.


\bi{epw1939}
Wigner, E. P., {\it Ann. Math.\/} {\bf 40}, 149 (1939).

\bi{lhr}
Ryder, L. H., {\it Quantum Field Theory,\/} Cambridge: Cambridge 
University Press, 1996.

\bi{gs}
Sterman, G., {\it An Introduction to Quantum Field Theory,\/}
 Cambridge: Cambridge 
University Press, 1993.

\bi{sw}
Weinberg, S. {\it The Quantum Theory of Fields, Vol. I\/}
 Cambridge: Cambridge 
University Press, 1995.

\bi{r1}
Ahluwalia, D. V. {\it Found. of Phys.\/} {\bf 28}, 527 (1998).


\bi{r2}
Ahluwalia, D. V. {\it Found. of Phys. Lett.\/} {\bf 10}, 301 (1997).



\bi{gg}
Gaioli, F. H., and Garcia Alvarez, E. T.,
{\it Am. J. Phys.\/} {\bf 63}, 177 (1995).

\bi{rm}
Mohapatra, R. N., {\it Massive Neutrinos in Physics and Astrophysics\/},
Singapore: World Scientific, 1991.


\bi{dva1996}
Ahluwalia, D. V., {\it Int. J. Mod. Phys. A\/}
{\bf 11}, 1855 (1996).

\bi{agj}
Ahluwalia, D. V., Johnson, M. B., and Goldman, T.,
{\it Mod. Phys. Lett. A\/} {\bf 9}, 439 (1994).

\bi{Hladik}
Hladik, J., {\it Spinors in Physics\/},
New York: Springer, 1999.

\bi{f}
Funakubo, K., Kakuto, A., Otsuki, S., and Toyoda, F.
{\it Prog. Theor. Phys.\/} {\bf 95}, 929 (1996)

\bi{cp}
Ahluwalia, D. V. {\it Mod. Phys. Lett. A\/} {\bf 13}, 3123 (1998).


\bi{vd1}
Dvoeglazov, V. V., {\it Nuovo Cim A\/} {\bf 108}, 1467 (1995)


\bi{vd2}
Dvoeglazov, V. V., {\it Int. J. Theor. Phys.\/} {\bf 34}, 2467 (1995).

\bi{grf1994}
Ahluwalia, D. V., {\it Phys. Lett. B\/} {\bf 339}, 301 (1994).

\bi{pla2000}
Ahluwalia, D. V., 
{\it Phys. Lett. A\/} {\bf 275}, 31 (2000).

\bi{ak}
A. Kempf, G.  Mangano,  R. B.  Mann, 
Phys. Rev. D {\bf 52} 1108,  (1995).

\bi{sdh}
S. de Haro, 
Class. Quantum Grav. {\bf 15}, 519 (1998).


\bi{as}
R. J. Adler, D. I. Santiago, 
Mod. Phys. Lett. A {\bf 14}, 1371 (1999).

\bi{gac}
G. Amelino-Camelia, 
Mod. Phys. Lett. A {\bf 12}, 1387 (1997).

\bi{mm}
M. Maggiore, 
Phys. Lett. B {\bf 304}, 65 (1993).


\bi{ns}
N. Sasakura, 
Prog. Theor. Phys.  {\bf 102}, 169 (1999).

\bi{fs}
F. Scardigli,
Phys. Lett. B. {\bf 452}, 39  (1999).

\bi{cls}
S. Capozziello, G.  Lambiase, G. Scarpetta, 
Int. J. Theor. Phys. {\bf 39}, 15 (2000).

\bibitem{jm}
J. Madore, gr-qc/9709002.

\bibitem{ac}
A. Connes, 
{\it  Noncommutative Geometry,\/} San Diego: 
Academic Press,  1994.

\bi{sm}
S. Majid, q-alg/9701001.

\bi{string_ur}
G. Veneziano, 
Europhys. Lett. 2 (1986) 199.

\bi{cr1}
Olinto, A. V. {\it Phys. Rept.\/} {\bf 333-334}, 329 (2000)

\bi{cr2}
Coleman, S., and  Glashow, S. L., 
{\it Phys. rev. D\/}, {\bf 59}, 116008 (1999).

\bi{cr3}
Amelino-Camelia, G, and Piran, T.,
astro-ph/0008107.
q
\bi{cr4}
Bertolami, O., {\it Nucl. Phys. Proc. Suppl.\/} {\bf 88}, 49 (2000).

\bi{cr5}
Bertou, X., Boratav, M., Letessier-Selveon, A. 
{\it Int. J. Mod. Phys. A\/} {\bf 15}, 2181 (2000).  

\bibitem{yong}
Ahluwalia, D. V., Liu, Y., and Stancu, I., hep-ph/0008303.

\end{references}
\end{document}